# Improving Latency in a Signal Processing System on the Epiphany Architecture

Peter Brauer
Ericsson AB
Goteborg, Sweden
peter.brauer@ericsson.com

Martin Lundqvist
Ericsson AB
Goteborg, Sweden
martin.lundqvist@ericsson.com

Aare Mällo
Ericsson AB
Goteborg, Sweden
aare.mallo@ericsson.com

*Abstract* - **In this paper we use the Adapteva Epiphany manycore chip to demonstrate how the throughput and the latency of a baseband signal processing chain, typically found in LTE or WiFi, can be optimized by a combination of task- and data parallelization, and data pipelining. The parallelization and data pipelining are facilitated by the shared memory architecture of the Epiphany, and the fact that a processor on one core can write directly into the memory of any other core on the chip.**

## 1. Introduction

Baseband signal processing in wireless systems such as WiFi and LTE are often implemented in ASIC's, FPGA's, and DSP's. During the last decade various alternative, manycore architectures for this domain have emerged (e.g. [1], [2], and [3]). One recent example is the Epiphany chip from Adapteva [4]. The Epiphany architecture offers several attractive features, e.g. scalability up to thousands of cores, high energy efficiency, floating-point support, and also a straightforward programming interface in ANSI C and a gcc based compiler.

In the present paper we do a hands-on evaluation on the 64-core version of the Epiphany chip and implement part of an OFDM (Orthogonal Frequency Division Multiplexing [5]) based signal processing chain (IFFT, deinterleaving, and demapping).

The goal of the evaluation is to see how well the Epiphany architecture is suited for this type of signal processing, and especially to explore different techniques for increasing the throughput and reducing the latency.

Programming an embedded many-core system becomes more complex with increasing number of CPU's. Domain-specific tools and languages designed to reduce this problem have been proposed, e.g. the StreamIt [6], and the ΣC [7] languages. However, for the Epiphany chip no such tools yet exist, and we have therefore been forced to use handwritten code for all parts of the problem.

The outline of the paper is as follows. The Epiphany architecture is described in section 2, and in section 3 the application is presented. In section 4 three alternative implementation cases of the signal processing chain are presented. Case I is a single core reference implementation, case II is a multicore implementation where task parallelization is used to enhance the throughput, and case III is a multicore implementation where data parallelization and data pipelining are employed to further reduce latency. Implementation details and measurement results for the three cases are given in section 5. Finally, in section 6 we summarize our conclusions and in section 7 we discuss some findings and opinions concerning improvements in software design for manycore deployment.

## 2. Epiphany Architecture

The Epiphany is a 2D mesh manycore architecture, with distributed shared memory, as shown in Figure 1. The current architecture supports up to 4096 cores, each consisting of a RISC CPU, two DMA engines, a local memory of 32 kB, and a network interface.

The RISC CPU ISA is optimized for real-time signal processing and contains 35 instructions. Each CPU can execute two floating point operations (i.e. one floating point MAC) and a 64-bit memory load operation on every clock cycle. This totals an on-chip processing availability of over 75 GFLOPS/W (CPUs running at 600 MHz) for the Epiphany-IV used for our evaluations.

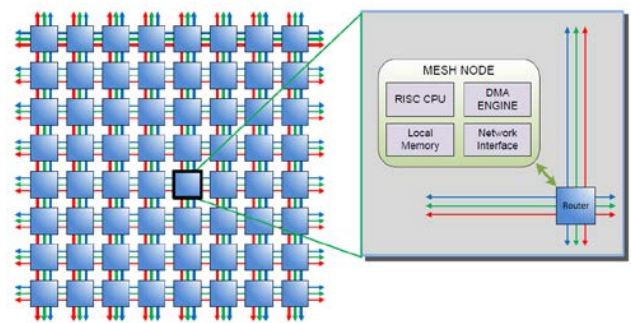

**Figure 1 - Adapteva's Epiphany architecture**

The local memory is intended for both code and scratch pad data, and it is directly addressed by its unique 32-bit address from anywhere on the chip, from any of the CPUs or via a DMA engine.

The router is the building block of the Epiphany Network-on-Chip (eMesh). Three separate networks are handled by the router, one for on-chip write traffic, one for on-chip read traffic, and one for off-chip read and write traffic. The "on-chip write" (called cMesh) is the fastest of these three networks and has a maximum bidirectional throughput of 8 bytes/cycle in each of the four routing directions with zero start-up time.

## 3. Wireless receiver signal processing

The application used for the evaluations in this paper is a part of a typical wireless baseband receiver signal processing system, and is shown in Figure 2. It contains three tasks: IFFT, deinterleaving, and demapping. A description of these processing tasks, as well as source code, can be found in [8].

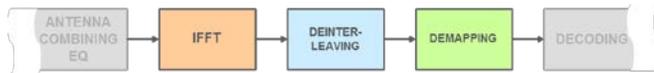

**Figure 2 – Wireless receiver signal processing chain**

In such a wireless system, the radio is continuously streaming data to the baseband processing unit at a certain rate, defining the throughput requirements of the system.

Also, in most wireless systems, the receiver is supposed to send a response within a certain time – based on the contents of the data – defining the latency requirements of the system.

As an example of typical requirements we refer to the 3GPP LTE system [9], able to handle 12 OFDM symbols (12 to 1200 complex values) in 1 ms, or 83 µs per symbol. An exact requirement on the latency of each individual task in the signal processing chain is difficult to state, since the requirement applies to the system as a whole. However, a rough estimate is that in an LTE system the total latency of the three tasks in Figure 2 should not be larger than 100 µs.

## 4. Optimization approach

We introduce optimizations gradually, in order to gain either throughput or latency. Analyzing the properties of our signal processing chain, and taking advantage of the Epiphany architecture, we utilize parallelism, shared memory communication, data locality, and geographical considerations.

The throughput and latency optimizations are illustrated below in the form of three cases.

Case I is a reference implementation where all tasks are executed on a single core.

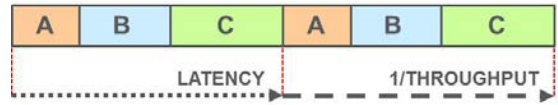

**Figure 3 - Case I: Single core deployment**

Case II utilizes task parallelization over multiple cores, and case III introduces data parallelization and data pipelining.

### 4.1 Single core deployment

In case I our signal processing chain is deployed on a single core, as shown in Figure 3.

The latency is the sum of the execution time of the individual tasks in the signal processing chain, and the throughput is limited by the latency of the total signal processing chain.

The cost of data movement and synchronization in this case is zero, since all processing tasks use the same physical memory, and every task starts immediately once its input data is provided as output data from the previous task.

### 4.2 Task parallelization

In case II task parallelization is introduced by deploying each task on separate cores, as shown in Figure 4, and synchronizing them in a data driven fashion.

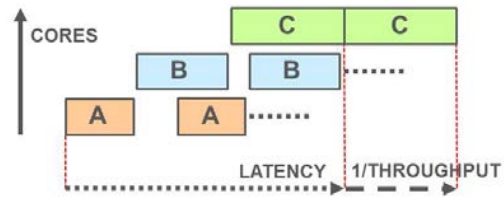

**Figure 4 - Case II: Increasing throughput using task parallelization**

The throughput is now limited only by the processing time of the most time consuming task – a significant improvement compared to the single core deployment.

The latency is similar to the previous case, except for the cost of synchronizing the start of the next task on another core.

There is no latency increase for a task continuously writing its output data to the memory of another core on the Epiphany, as shown in Figure 5, only a delay for the data to actually travel over the mesh to the other core.

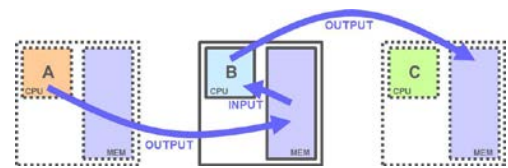

**Figure 5 - Reading data locally, writing data remotely**

### 4.3 Data parallelization and pipelining

In case III data parallelization and data pipelining are introduced, in order to reduce latency. Data parallelization may be applied to all three tasks in our processing chain, whereas data pipelining is only possible between the deinterleaving and demapping tasks, due to the IFFT task producing its output data in complete blocks, rather than in a stream.

The purpose of this case is primarily to propose and evaluate different methods for reducing latency of a signal processing chain on the Epiphany, rather than just chasing the minimum value. We have therefore chosen to parallelize only the IFFT, and to utilize data pipelining between the demapping and the deinterleaving, as shown in Figure 6.

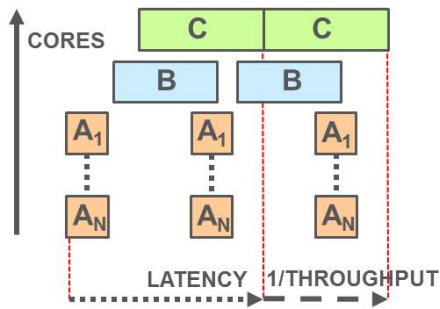

**Figure 6 - Case III: Reducing latency using data parallelization and pipelining**

A parallelized IFFT needs to continuously interchange data between its parallelized tasks, something efficiently facilitated on the Epiphany by deploying them on neighboring cores.

Data pipelining in the interface between the deinterleaving task and the demapping task is introduced by reducing the size of data written by the former before synchronizing a read of the latter, increasing task concurrency.

The throughput in this case should be roughly the same as in case II, while the latency may be substantially decreased, limited mainly by the most time consuming of the tasks.

For manycore deployments of complete signal processing systems, there will always be tradeoffs to be considered, settling for sufficient real time gains while not exaggerating resource consumptions. This optimization balance is discussed more in detail in the next section.

### 5. Implementation and results

The throughput and the latency for the different cases have been measured on the Parallella board [10], and the OFDM size used is 256 subcarriers.

A summary of the measurements is given in Table 1 below. Boldface numbers indicate tasks that determine the total throughput for each case.

**Table 1 - Latencies and Throughput for cases I, II, III.**

|  | I | II | III |
|---|---|---|---|
| IFFT [cycles] | 18862 | 18863 | 2958 |
| Deinterleaving [cycles] | 45043 | 45046 | 47585 |
| Demapping [cycles] | 46377 | **46377** |  |
| Total latency [cycles] | **110282** | 110286 | 50543 |
| Throughput [symbols/s] | 5441 | 12937 | 12609 |

The system clock on the Epiphany chip is 600 MHz, and our measurements results have a resolution proportional to this. The individual timers on each core are hard to synchronize perfectly, and this synchronization deviation – a couple of cycles per core - even harder to verify. Every measurement also takes a certain amount of time to perform, albeit less than 100 cycles.

In Table 1 these measurement times have been subtracted, and since the effects of our optimizations are quite large and clearly visible, we feel confident that remaining uncertainties do not affect our conclusions.

In the following subsections each case is discussed in more detail.

### 5.1 Case I

Case I is included here as a reference implementation. The three tasks are executed sequentially on a single core. Measures for one OFDM symbol gives 110282 cycles, which corresponds to a data rate of one symbol per 182µs. LTE uses a data symbol rate of 83µs, and the obvious conclusion is that the deployment in case I cannot meet the throughput requirement.

### 5.2 Case II

In case II the three tasks are distributed over three cores, and executed in parallel. Data is synchronized in blocks of 256 samples between the cores. The throughput is determined by the task with the largest latency (demapping), consuming 46377 cycles, or 77 µs. This deployment is able to handle LTE data rates. The total latency of the processing chain is just a few cycles higher than in case I, 110286 cycles, due to overhead in synchronization between cores.

**Synchronization and barriers**

Our platform offers no means of implicit synchronization – as an ASIC or FPGA deployment would – so when utilizing more than one core, data processing on different cores needs to be synchronized in software. This synchronization will introduce a total processing time overhead, which must be kept as small as possible not to ruin the multicore speedup. Also, any buffer over-run or dead-lock situation needs to be avoided by designing a correct and application specific implementation.

In C-syntax this simple synchronization code can be written as:

Producer:
```
*remote_flag_p = 1;
```
Consumer:
```
while (*new_data_flag_p != 1);
*new_data_flag_p = -1;
```

The pointers refer to the same flag, located in the local memory of the data consuming task. Analyzing the generated assembler code, this synchronization overhead is 1 clock cycle in the producer task and at a maximum 6 clock cycles in the consumer task. Barriers joining several producer tasks use separate flags for each one, sequentially polled in the consumer task.

### 5.3 Case III

In case III we try to reduce the total latency by a combination of data parallelization and more fine-grained data synchronization between the cores. Below we describe how the degree of parallelization and the granularity of this data synchronization have been determined.

**IFFT Parallelization**

When implemented on a single core the IFFT consumes slightly less than 19 000 CPU cycles. In separate tests we have measured the IFFT running on 1, 2, 4, 8, 16 and 32 cores. The test results are shown in Figure 7 where the IFFT latency is plotted as a function of the number of cores, $N_C$.

With increasing $N_C$ the latency first decreases as expected, but then actually increases for $N_C > 16$. This effect is due to the cost of data synchronization, which is proportional to $N_C$. As $N_C$ increases, the synchronization cost eventually grows larger than the cost of the mathematical operations in the FFT itself.

As a trade-off between latency gain versus resource consumption we have chosen to use $N_C = 8$ parallel cores for the IFFT in this implementation. The latency for the IFFT is then reduced from 18863 to 2958 cycles according to Table 1, corresponding to a parallelization efficiency of 18863 / (8 × 2958), or roughly 80 %.

**Fine-Grained Data Pipelining**

The algorithms within the deinterleaving and demapping tasks are able to process data on a sample basis. By inserting synchronization code into the functions, data can be synchronized with minimum overhead. The optimum data block size must be determined, and this is dependent of the relation between synchronization overhead and actual processing. Figure 8 shows the total latency of the combined deinterleaving and demapping tasks as a function of data block size.

In this case, the lowest latency is achieved with 1 sample synchronization between the tasks.

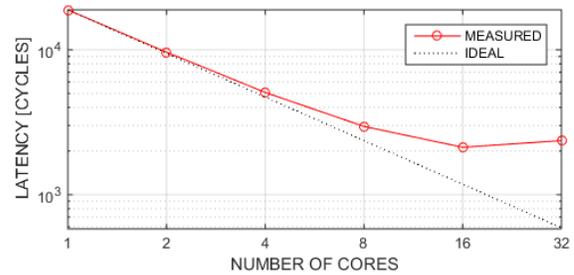

Figure 7 – Parallel IFFT: Latency vs number of cores

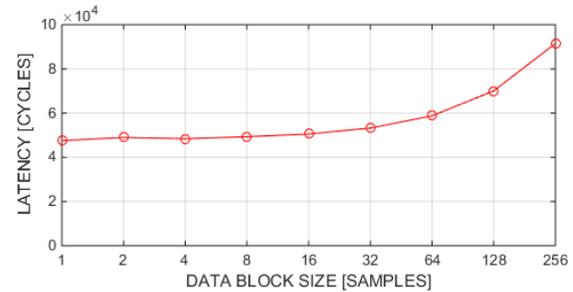

Figure 8 - Combined deinterleaving-demapping: Latency versus data block size.

**Putting It All Together**

In Figure 9, the throughput and the latency for all cases in Table 1 are compared in two diagrams. The upper diagram shows that the total throughput is more than doubled when task parallelization is introduced in case II, and the lower diagram shows that the latency is more than halved when data parallelization and fine-grained data pipelining is introduced in case III.

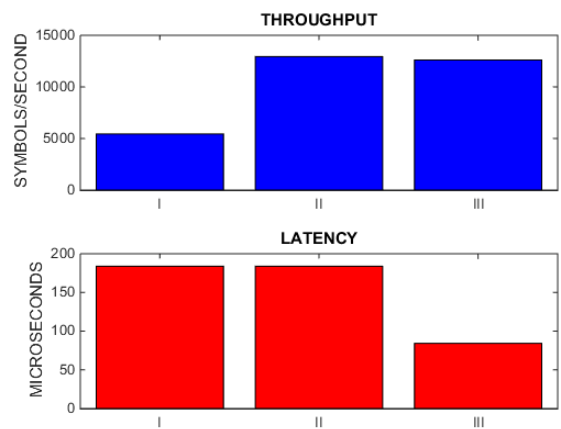

**Figure 9 - Throughput and latency for cases I, II, and III. Case III achieves both high throughput and low latency by a combination of task- and data parallelization, and fine-grained data pipelining.**

## 6. Conclusions

By using a combination of task and data parallelization and fine-grained data pipelining, we have been able to decrease the processing latency of a typical signal processing chain with more than 50%. Parallelization requires more resources while fine-grained data pipelining can be accomplished by customizing interfaces, but without additional CPU's.

The degree of parallelization and the granularity in the data pipelining are design parameters that must be determined by measurements on the actual hardware. The need for these types of measurements is clearly illustrated by the IFFT parallelization in section 5.3.

Our evaluations indicate that the Epiphany architecture, with its communication meshes and shared memory space, is suitable for deployments of such data streaming signal processing system with only a modest overhead.

## 7. Discussion

Manycore hardware is certainly here to stay and the Epiphany architecture has several useful features - most notably its scalability, its simplicity, and its shared memory architecture.

In the present evaluation we have only considered a limited part of a signal processing system, requiring around 10 cores. We have manually written C-code for all tasks and signal processing as well as for all platform related functionality. However, for a complete and realistic signal processing system comprising hundreds or thousands of cores this approach would not work.

In signal processing design it is desirable to develop the specific algorithms independently from the actual deployment. For the Epiphany as well as for many embedded systems the only available implementation language with compiler support is C or sometimes C++. Therefore, the implemented functions are traditionally written and optimized to work on arrays of data.

In a streaming-like processing context, data buffer sizes must be handled in respect to the deployment, and synchronization primitives have to be intertwined with the signal processing code. When coding platform functionality using C you have total control of these details, at the same time as letting you introduce dangerously bold approaches to distributed processing – in this sense C is a blessing as well as a curse!

We definitely see a need for separation of concerns, regarding the responsibilities of the application and the platform when deploying an efficient signal processing system for manycore. It's not reasonable for designers to have expert knowledge in every domain, ranging from application algorithms to hardware architecture details. Neither should we expect automatic tools to magically adapt and optimize any legacy implementation to take advantage of every processing resource available.

A reasonable vision to strive for may instead be the use of application domain languages, where designers could identify and express properties suitable for manycore compilers to exploit, without actually considering particular hardware details.

Suggestions on this theme are discussed in different forums, for example in [11], and interesting research to combine higher abstraction languages with data flow is ongoing, one example is enabling stream processing using Feldspar [12].

Needed for the Epiphany architecture is an execution platform, reducing the need for manual labor, and automatically creating the code for task execution and data transport and synchronization.

## References

[1] http://www.kalrayinc.com/

[2] http://www.tilera.com/

[3] Howard, J. et al.: A 48-Core IA-32 message-passing processor in 45nm CMOS using on-die message passing and DVFS for performance and power scaling. IEEE J. of Solid-State Circuits, vol. 46, pp. 173-183, 2011.

[4] http://www.adapteva.com/

[5] Tse, D., and Viswanath, P., Fundamentals of Wireless Communication (2005)

[6] Thies, W. et al. StreamIt: A Compiler for Streaming Applications. In *Proceedings of the 11th International Conference on Compiler Construction* (CC '02), R. Nigel Horspool (Ed.). Springer-Verlag, London, UK, UK, 179-196, 2002.

[7] Goubier, T. et al. ΣC: A Programming Model and Language for Embedded Manycores. In *Algorithms and Architectures for parallel processing*, 385-394. Springer Berlin Heidelberg, 2011.

[8] Själander, M. et al, An LTE Uplink Receiver PHY Benchmark and Subframe-Based Power Management. *IEEE International Symposium on Performance Analysis of Systems and Software,* 2012.

[9] ETSI TS 136 201 V12.2.0 (2015-04)

[10] http://www.parallella.org/

[11] Hwu, W. M. et al. Implicitly parallel programming models for thousand-core microprocessors. In *Design Automation Conference, 2007. DAC'07. 44th ACM/IEEE*, 754-759. IEEE, 2007.

[12] Aronsson, M. et al. Stream Processing for Embedded Domain Specific Languages. In *Proceedings of the 26nd 2014 International Symposium on Implementation and Application of Functional Languages (IFL '14),* 2014.